\documentstyle[11pt,aaspp4,flushrt,psfig,graphicx]{article}

\def\etal{\it et al. \rm }
\begin{document}

\title{Age and Metallicity Estimation of Globular Clusters from Str\"omgren Photometry}

\author{Karl Rakos}
\affil{Institute of Astronomy, Vienna University, Vienna, Austria; karl.rakos@chello.at}
\author{James Schombert}
\affil{Department of Physics, University of Oregon, Eugene, OR; js@abyss.uoregon.edu}

\begin{abstract} 

We present a new technique for the determination of age and metallicity in composite stellar
populations using Str\"omgren filters.  Using principal component (PC) analysis on multi-color
models, we isolate the range of values necessary to uniquely determine age and metallicity
effects.  The technique presented herein can only be applied to old ($\tau >$ 3 Gyrs) stellar
systems composed of simple stellar populations, such as globular clusters and elliptical
galaxies.  Calibration using new photometry of 40 globular clusters with spectroscopic [Fe/H]
values and main sequence fitted ages links the PC values to the Str\"omgren colors for an
accuracy of 0.2 dex in metallicity and 0.5 Gyrs in age.

\end{abstract}

\section{INTRODUCTION}

The two primary processes that determine the characteristics of a stellar population are its
star formation history and its chemical evolution.  For an actively star forming system, such
as the disk of our Galaxy, these two process are intertwined and will display a feedback loop as
star formation continues and, thus, understanding an active system requires detailed HR
diagrams and individual stellar spectroscopy.  However, a simple stellar population (SSP), one
formed in a single event from a single cloud of gas (e.g. a globular cluster), will have a
fixed metallicity and age that may be derived from the color-magnitude diagram (CMD).  A burst
stellar population, i.e. one derived from a extended star formation event, will be composed of
a combination of SSP's and the evolutionary processes are reflected into the population's age and
metallicity by the luminosity weighted mean of the various SSP's. It is possible to
characterize a burst population if the duration of the burst is short and distribution of
metallicities is uniform (Renzini \& Buzzoni 1986).

Early studies of composite stellar populations focused on broadband colors of spiral bulges and
elliptical galaxies (Sandage \& Visvanathan 1978, Tinsley 1980, Frogel 1985) and these datasets
supported the hypothesis that red galaxies are composed, primarily, of old, metal-rich stellar
populations under the burst hypothesis.  Unfortunately, it was quickly realized that detailed
interpretation of broadband colors with respect to age and metallicity are complicated by
several factors.  Foremost was the assumption that old stellar systems are composed of a
uniform population in age and metallicity.  It was soon demonstrated by population synthesis
techniques (O'Connell 1980) that a young population quickly reddens to similar integrated
colors as an old population and that the change in color is abrupt even while the differences
in age may be quite large (greater than 5 Gyrs).  It was also noted by Burstein (1985) who
found significant variations in the age and metallicity properties of galactic globular
clusters which did not reflect into their integrated broadband colors.  Lastly, it was
identified through the use of stellar population models that slight changes in age and
metallicity operate in the same direction of spectroevolutionary parameter space (Worthey
1994).  This coupling of age and metallicity (known as age-metallicity degeneracy, Worthey
1999) is due to competing contributions from main sequence turn-off stars (sensitive to age)
and red giant branch (RGB) stars (sensitive to metallicity) near 5000\AA.  Filters that bracket
this region of a galaxy's spectrum will require increasingly accurate values for metallicity to
determine a unique age and vice-versa.

To avoid the age-metallicity degeneracy problems, a majority of recent stellar population
studies have focused on the determination of age and metallicity through the use of various
spectral signatures, such as H$\beta$ for age (Kuntscher 1998, Trager \etal 2001).  This
approach provides a finer comparison to stellar population models, but requires assumptions
about the relationship between metallicity indicators (e.g.  Mg$_2$) and the [Fe/H] value of
the population as reflected into the behavior of the red giant branch.  In other words,
spectral lines provide the value of that element's abundance, but what is really required is
the temperature of the RGB which is a function of the total metallicity, $Z$.  Varying ratios
of individual elements to $Z$ complicates the interpretation of line studies (Ferreras, Charlot
\& Silk 1999).  In addition, these techniques have limitations due to the required high S/N for
the data that make them problematic for the study of high redshift systems.

An alternative approach to spectral line studies is to examine the shape of specific portions
of a spectral energy distribution (SED) using narrow band filters centered on regions sensitive
to the mean color of the RGB (metallicity) and the main sequence turnoff point (mean age)
without the overlap that degrades broadband colors.  The type of galaxy examined, for example
stellar systems with ongoing star formation where a mix of different age populations may be
present, will still limit this technique.  However, for systems that have exhausted their gas
supply many Gyrs ago (i.e. old and quiescent), it may be possible to resolve the underlying
population with some simple assumptions on their star formation history and subsequent chemical
evolution.  Thus, we have the expectation, guided by the results of evolution models, that
objects composed of SSPs or a composite of SSPs (e.g. ellipticals) present special
circumstances where the age-metallicity degeneracy can be resolved and allow the study of the
evolution of stellar populations.

In a series of earlier papers, we have examined the $uz,vz,bz,yz$ colors of globular clusters
and used a combination of their colors and SED models to derive the mean age and metallicity of
dwarf, bright and field ellipticals (Rakos \etal 2001, Odell, Schombert \& Rakos 2002, Rakos \&
Schombert 2004).  While spectroscopic data is superior for age and metallicity estimations in
high S/N datasets, our goal has been to develop a photometric system that can be used for
galaxies of low surface brightness and/or high redshift, where spectroscopy is impractical or
impossible.  Our past technique has been to relate the $vz-yz$ color index to mean metallicity,
since the $vz$ filter is centered on the absorption line region near 4100\AA\, as guided by our
multi-metallicity SED models.  The $bz-yz$ color index, whose filters are centered on continuum
regions of the spectral energy curve, measures the mean stellar age.  This method was crude
since metallicity changes will move the effective temperature of the RGB and, thus, the
continuum $bz-yz$ colors.  In our more recent work, we have included photometry through the
$uz$ filter which provides an additional handle on age and metallicity effects in the other two
color indices.  In addition, we have applied a principal component (PC) analysis on the
multi-color data (Steindling, Brosch \& Rakos 2001) that more fully isolates metallicity from
age effects and the changes due to recent star formation.

The success of our narrow band, multi-color technique has motivated us to return to our
original calibration objects, i.e. globular clusters, and obtain higher accuracy photometry and
investigate their color behavior under PC analysis.  In addition, new globular cluster age
estimates are available in the literature (see Salaris \& Weiss 2002), based on direct
determination from the main sequence turnoff point rather than SED models, and Schulz \etal
(2002) have published a set of new SED models for metallicities from $-$1.7 to $+$0.4 and ages
from 10$^6$ yrs to 16 Gyrs using the most recent isochrones from the Padova group.  Our aim for
this paper is threefold: 1) to present our new photometry based on the integrated light from 40
Milky Way globular clusters with well determined ages and metallicities (baseline SSP's), 2) to
demonstrate that narrow band filters are effective in discriminating age and metallicity for a
single generation objects by comparison to data and SED models and 3) calibrate our photometric
system to globular cluster ages and metallicities and explore its use, and limitations, as an
age and metallicity estimator for dwarf and giant ellipticals.

\section{DATA AND MODELS}

Our primary data sample is the photometry of Milky Way globular clusters published in Rakos
\etal (2001), with a few corrections, plus new observations obtained during the 2003 CTIO
observing season.  There are 40 clusters in the final sample where a small number of clusters
were observed multiple times.  The photometric accuracy is on the order of $\pm$0.02 mag in
$bz-yz$ and $vz-yz$ plus $\pm$0.03 mag in $uz-vz$.  The Str\"omgren filters have a long history
in the literature of be used to determine metallicity, surface gravity and effective
temperature in stars (Bell \& Gustafsson 1978), but their use for a composite system, one
composed of many stars of different luminosities and temperatures but a single metallicity, is
more complicated.

\begin{figure}
\centering
\includegraphics[width=12cm,angle=-90]{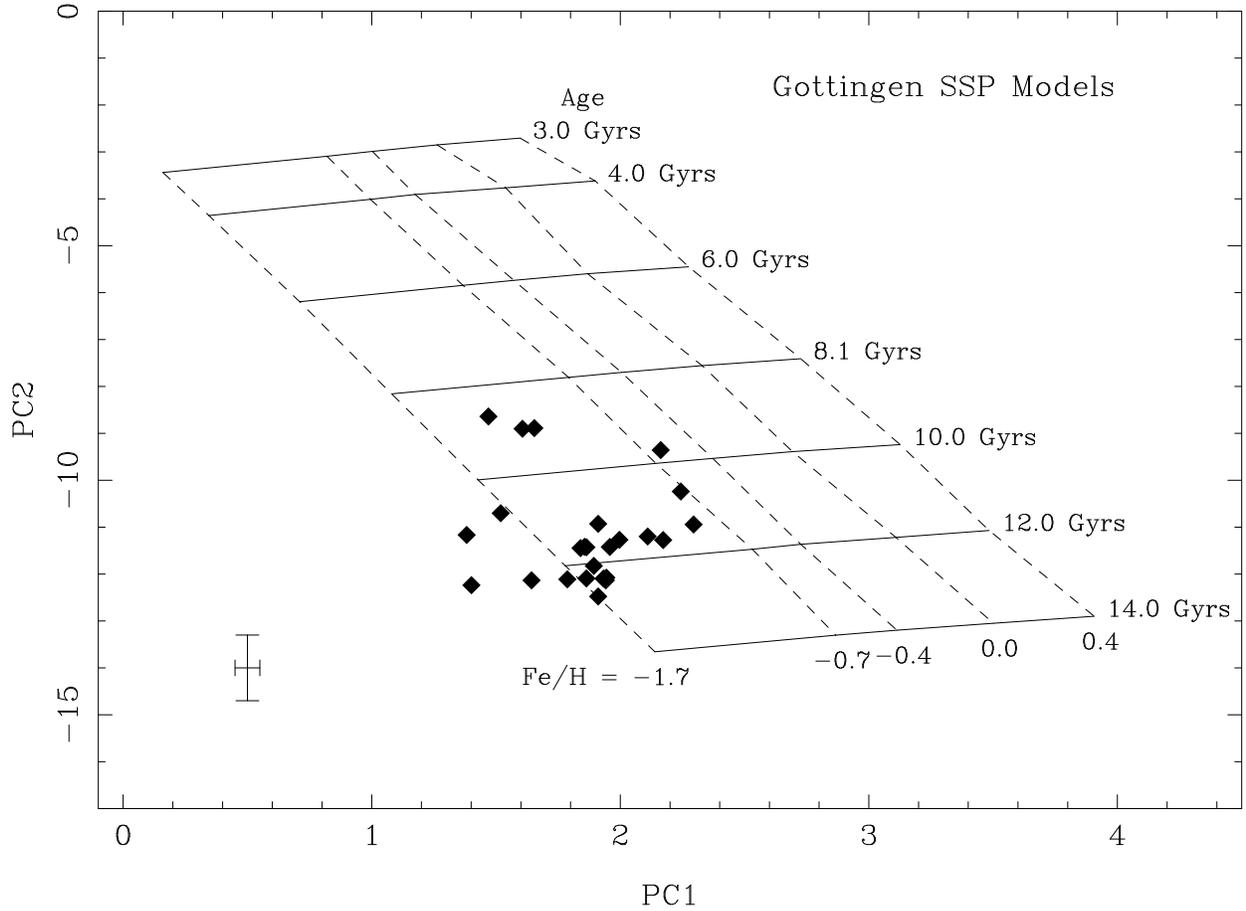}
\caption{The principal component axes PC1 versus PC2 for the Schulz \etal SED models.  Each
intersection represents a single model.  Age runs roughly vertical to PC1, [Fe/H] runs
horizontal.  Globular clusters with reliable ages, metallicities and narrow band colors are
shown are solid symbols (ages and metallicities from Salaris \& Weiss).  The error bar in the
lower left-hand corner display the typical error for the Salaris \& Weiss data converted to PC
axis.  They occupy the proper region of the grid indicating that the SED models correctly
describe their color indices.}
\end{figure}

Our first challenge is to demonstrate that age and metallicity in a stellar population can be
discriminated solely from its narrow band colors.  To this end, we first need to track the
behavior of our colors by convolving our filters to a set of SED models with varying age and
metallicities.  Our choice of SED models is the recent work published by Schulz \etal (2002)
which provide $uvby$ (rest frame $uz,vz,bz,yz$) color indices for SSP models using five different
metallicities ($-$1.7, $-$0.7, $-$0.4, 0.0, $+$0.4) and ages between 0.8 to 14 Gyr.  We have
transformed these published indices to our modified system using the transformations shown in
Rakos, Maindl \& Schombert (1996) and applied the same PC analysis to the resulting colors as
outlined in Steindling, Brosch \& Rakos (2001).

Scientific data in the astronomy are usually neither linear nor orthogonal at the same time.
Often, for small regions of data, linearity and orthogonality can be reached through the use of
PC analysis (Murtagh \& Heck 1987).  Close inspection of Schulz \etal theoretical models has
shown acceptable linearity for ages larger than 3 Gyrs over the full range of model
metallicities.  In this restricted region, it is possible to apply PC analysis to; 1) separate
the age and the metallicity of a stellar population, 2) select the most correlated variables
and 3) determine linear combinations of variables for extrapolation.  In the Table 1, we list
the metallicities, ages and corresponding photometric color indices ($uz-vz$, $bz-yz$, $vz-yz$)
for the range of models considered herein.  These correlations deliver principal components
PC1, PC2 and PC3 (see Steindling, Brosch \& Rakos 2001), but only first two components, PC1 and
PC2, have significant content for our purpose.  Their values are given by:

\begin{equation}
PC1 = 0.471[Fe/H] + 0.175\tau +0.480(uz-vz) + 0.506(bz-yz) + 0.511(vz-yz)
\end{equation}

\begin{equation}
PC2 = 0.345[Fe/H] - 0.935\tau + 0.047(uz-vz) - 0.061(bz-yz) + 0.020(vz-yz)
\end{equation}

\noindent where $\tau$ is the age of the stellar population in Gyrs.  Metallicity is found to
have similar weight in both equations, contrary to the age index that has a small weight in
the first equation and a large weight in the second expression.  Of course, knowledge of the
values for PC1 and PC2, combined with the photometric indices, would provide simple solutions
to the equations for age and metallicity.

For the models in Table 1, PC1 and PC2 are calculated from equations (1) and (2) where the
colors are given by the models and, of course, [Fe/H] and age are the inputs for each model.
Figure 1 displays PC1 versus PC2 for a range of models and demonstrates their orthogonality
between age and metallicity.  We only have PC1 and PC2 model values for a few discrete values
of metallicity and age, but a simple linear interpolation between these values produces PCs
with significant accuracy over the entire surface in Figure 1.  The advantage of the smooth
behavior for age and metallicity in the PC plane is that if one has all three narrow band colors
($uz-vz$, $vz-yz$ and $bz-yz$) then knowledge of the correct PC1 and PC2 values allows for the
unique determination of age and metallicity from equations (1) and (2).  If the PC values are
unknown, then an iterative search scheme could select a range of values for $\tau$ and [Fe/H],
determine PC1 and PC2 from Figure 1, then look how well those values and the observed colors
solve equations (1) and (2).  We will outline this technique in greater detail in the next
section.

For comparison of the models to main-sequence fitted ages and metallicities of globular
clusters, we have use the results from Harris (1996, updated in February 2003) and Salaris \&
Weiss (2002).  Salaris \& Weiss provide two additional estimates for metallicity with the
designation CG97 (Carretta \& Gratton) and ZW84 (Zinn \& West) as well as two additional
estimates for ages with the same designation.  We have used a mean value of all three values for
the age and metallicity of each cluster (the mean error from Salaris \& Weiss is $\pm$0.1 Gyrs
in age and $\pm$0.1 dex in [Fe/H]) and calculated PC1 and PC2 using the observed narrow band
colors and the fitted metallicities and ages.  The resulting PC values are shown in Figure 1 (solid
symbols) with an error bar shown in the bottom left-hand corner that represents the mean error
of the Salaris \& Weiss sample.  The globular cluster data occupies the correct portion of the
PC1-PC2 diagram and confirms that the SED models do, in fact, describe the observed colors of
the globular clusters.  This is not an obvious result, for previous SED models failed to match
the $uv$ colors of globular clusters by between 0.5 and 0.7 mags (Rakos \etal 2001).  It
appears that recent changes in overshoot calculations, important for the determination of
colors for the blue HB stars, have converged the model colors and the observed $uz-vz$ colors
of globular clusters.

\section{ITERATIVE PC SOLUTIONS}

Reproducing PC values for objects with known metallicities and ages only confirms the validity
of the Schulz \etal SED models.  A more powerful approach would be to derive metallicity
and age values using only the observed narrow band colors and knowledge of the behavior of the
PC surface (i.e. guided by the SED models).  This process begins with the production of a mesh
of PC1 and PC2 values, with the expectation that one pair in the computed mesh represents the
true value for an unknown age and metallicity.  To identify the true value of PC1, PC2 for an
object with observed colors, but unknown age and metallicity, we interpolate a pair PC1 and PC2
from the assumed age and metallicity (see Figure 1), use expressions (1) and (2) and fill in
the observed color indices plus the assumed age and metallicity and compare the calculated
pairs with the pairs derived from the colors.

In practice, the search for the correct age and metallicity is evaluated by calculating the
differences between the PC values determined from models (i.e. Figure 1) and those PC values
determined from equations (1) and (2) where the observed colors are the input.  The quality of
the solution is measured by the root mean difference (RMS) between the PC values over two
iteration loops, one for age and one for metallicity.  This computation consists of starting
with a value for age and increasing the value with an index of 0.1 Gyr.  Each age iteration has
a nested metallicity loop increasing in steps of 0.004 dex.  At each step, PC values are
interpolated from the Schulz \etal models for the step values of age and metallicity, then
compared to the PC values determined from expressions (1) and (2) again using the step values
of age and metallicity plus the observed colors.

To limit interpolation errors, the starting and ending values in the first loop are fixed on
values for the age of the models given by Schulz \etal with step values interpolated from
Figure 1.  Each step of the age loop has a second loop that iterates over the metallicity
interval.  At each step, metallicity is evaluated first using the age to calculate the
metallicity from equation (1), then age is calculated from equation (2) which is then returned
to equation (1), inserting the newly calculated value for age and calculating a better value
for the metallicity.

Each iteration produces a better approximation for age (using equation 2) and is repeated for
total of six values of age and metallicity.  In the case that the iterated pair of PC1 and PC2
represents the true value, we expect it to have the smallest deviation for the calculated set
of six ages and six metallicities.  Therefore, we determine the quality of the solution by the
sum of mean square deviations of ages and metallicities for the selected PC1 and PC2 values.
While this technique is not as rigorous as a least squares minimization between two functions,
it has the advantage of being a quantative measure of the fit and the RMS sum does become
smaller and smaller as we approach the proper values of age and metallicity.

Lastly, to avoid problems associated with small local minimums, we extend the search for at
least four or more values of age in steps of 0.1 Gyr that have the smallest sum of mean square
deviations (for a very limited range of metallicity), which implies the closest match to the
proper age and the metallicity for the system.  The resulting grid minimum that is searched for
should be deepest and widest within the whole range of PC1, PC2 values.  Unfortunately, in some
cases, there is a secondary minima produced by the errors for the observed color indices and
the best solution may have a broad range of age values with an equal quality of fitness.  A
detailed example of our procedure is given in the Appendix for globular cluster NGC 6397.

The above technique was applied to all 40 clusters with metallicity information available from
spectroscopy and age dating using CMD diagrams.  For one cluster, NGC 6218, no acceptable
solution was found, the rest are listed in Table 2 where $<$Fe/H$>$ and $<$$\tau$$>$ are the
values from Salaris \& Weiss and [Fe/H]$_{ph}$ and $\tau_{ph}$ the photometric determination of
metallicity and age.  A comparison of these two determinations are shown in Figures 2 and 3
where the solid blue lines represent a one-to-one relationship between $<$Fe/H$>$, $<$$\tau$$>$
and the photometrically determined [Fe/H]$_{ph}$, $\tau_{ph}$.  The solid black lines are a
least squares fit to the relationships.  Both relationships are well within the internal errors
for metallicity and age from main sequence fitting as will be discussed in the next section.

\begin{figure}
\centering
\includegraphics[width=16cm]{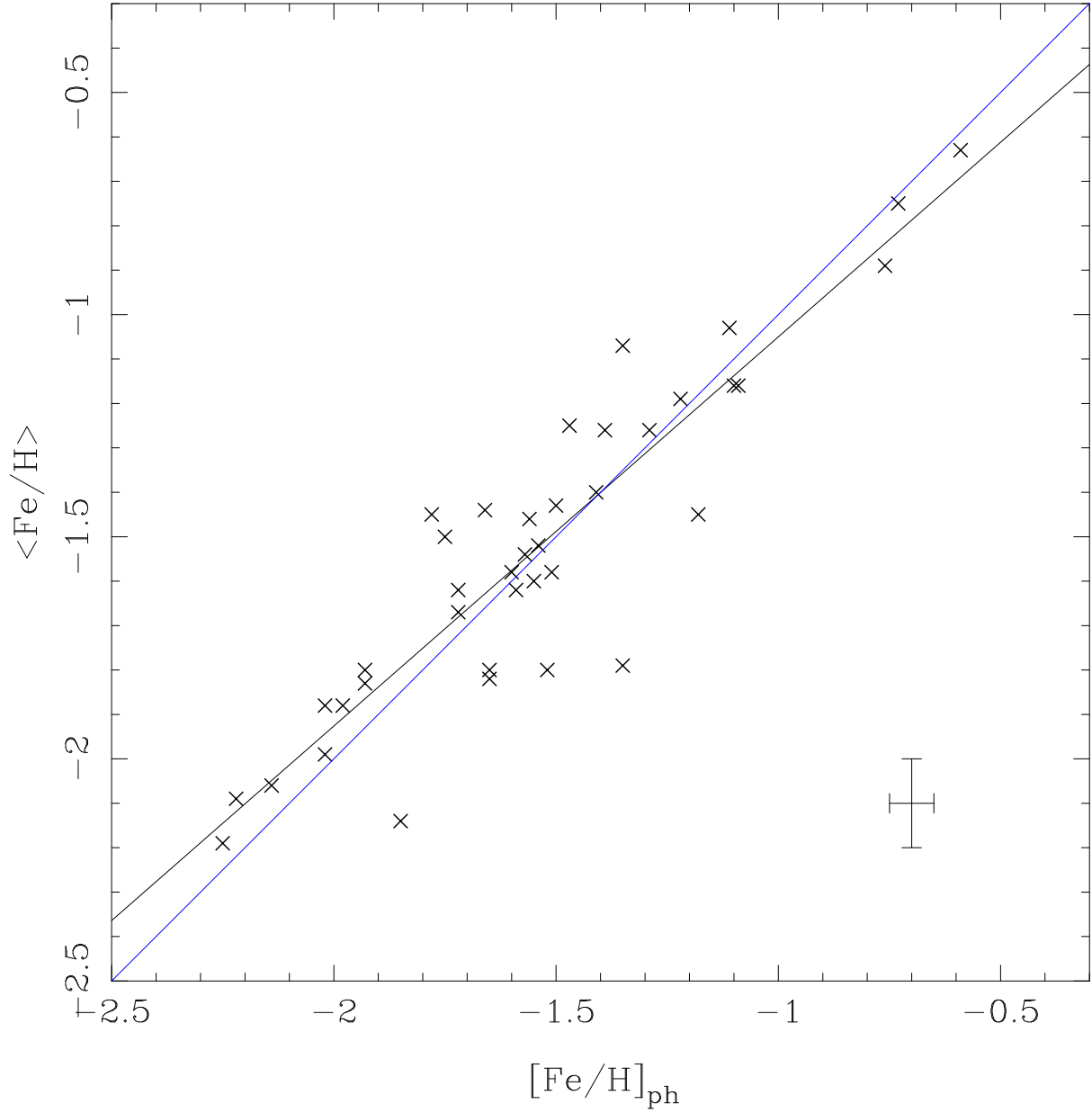}
\caption{The globular cluster metallicities as determined by the iterative technique
([Fe/H]$_{ph}$) versus their true metallicity ($<$Fe/H$>$, a mean value from the literature).
The blue line indicates a one-to-one relationship, the black line is a least square fit. The
error bar displays the typical uncertainty in metallicity from Salaris \& Weiss as well as the
uncertainly in our fitting technique.}
\end{figure}

\begin{figure}
\centering
\includegraphics[width=16cm]{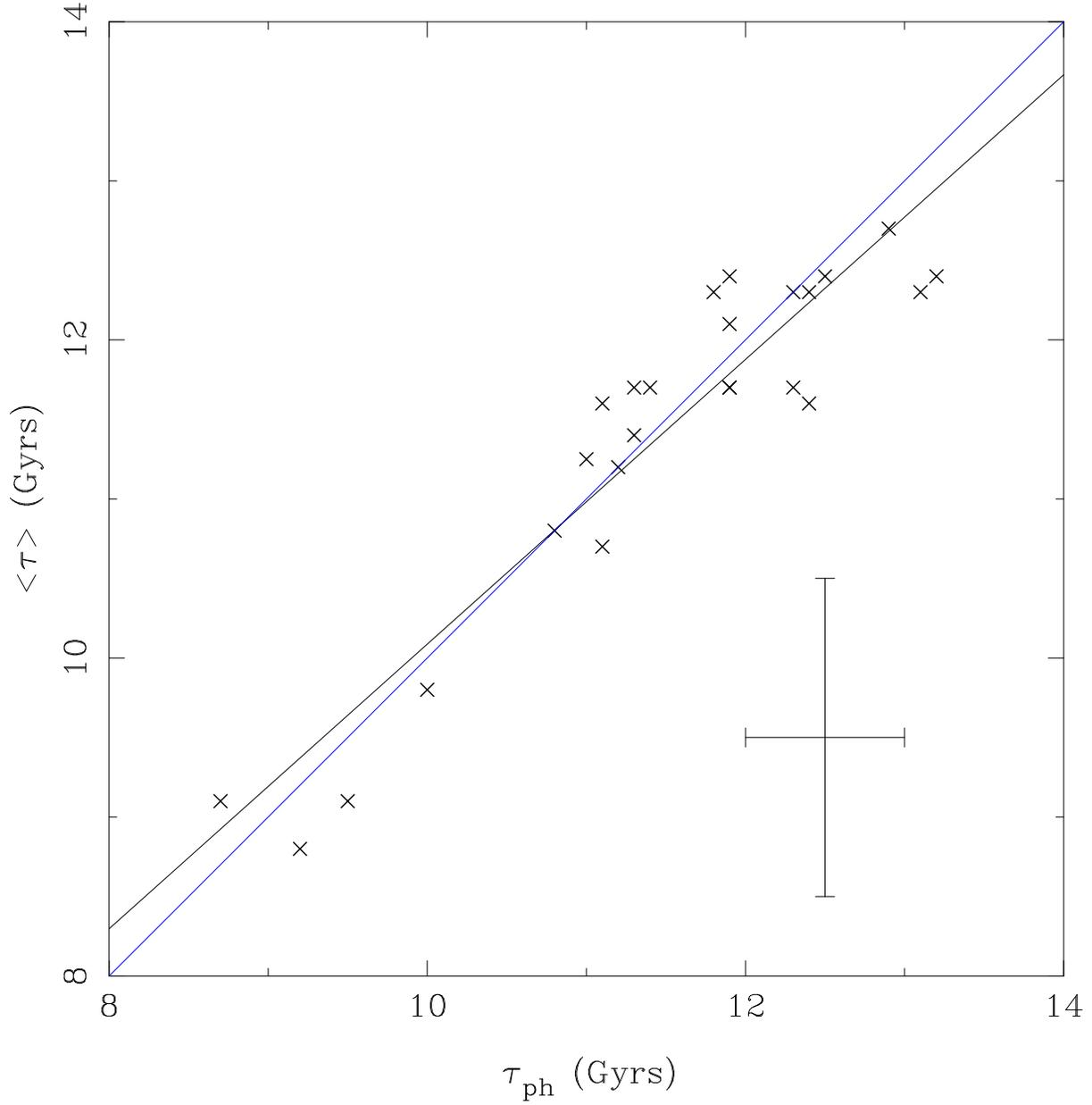}
\caption{The globular cluster ages as determined by the iterative technique ($\tau_{ph}$)
versus their true age from main sequence fitting ($<$$\tau$$>$, a mean value from the
literature).  The blue line indicates a one-to-one relationship, the black line is a least
square fit. The error bar displays the typical uncertainty in age from Salaris \& Weiss as well
as the uncertainly in our fitting technique. }
\end{figure}

\section{UNCERTAINTIES AND ERROR ESTIMATES}

The uncertainties to our technique fall into three categories; 1) observational
errors to the colors, 2) errors due to the quality of the solutions from the iterative method,
i.e. the range of equally likely solutions based on a quality of fit criteria and 3) the
ability of the Schulz \etal models to reproduce the underlying physical reality to the globular
cluster colors.  While we can directly test the impact of errors for the first two quantities,
we cannot address the meaning of the fits in terms of the various inputs to the population
models.  However, we can compare our technique results with the age and metallicities
determined by other means, in this case by comparing to the results from CMD fitting and
spectroscopy, as an indication of the merit of the models as applied to our method.

Observational error can arise through inaccuracy to the photometry either through simple
Poisson noise on the photon counts or internal errors to the standard stars or uncertain
Galactic reddening values.  As described in the previous sections, the data was obtained
under the best of conditions and the error is estimated to be 0.02 for $bz-yz$ and $vz-yz$ and
0.03 for $uz-vz$.  In order to estimate the effect of this error on the age and metallicity
determination, we have simply iterated the solution technique using a range of colors within
the measured errors.  Uncorrelated errors (i.e. randomly changing each color by its mean error)
produces solutions that varied by $\pm$0.5 Gyrs in age and $\pm$0.05 dex in [Fe/H].  Correlated
errors, for example moving all the colors to the blue, produced similar errors since bluer
$uv-vz$ colors increase age and metallicity fits whereas bluer $vz-yz$ and $bz-yz$ colors
decreased the fits.

Another source of uncertainty arises from the ability to determine a unique solution to the
given narrow band colors.  This, in effect, asks if the resulting PC pairs can produce a unique
set of age and metallicity values within a certain degree of accuracy.  One method of
estimating the quality of the solutions is to examine the behavior of the fits for a range of
age and metallicities near the chosen solution.  For our technique, this involves comparing the
RMS values for each of the nested loops.  The loops with the smallest deviations from the mean
are the ones closest to the derived solution (see the Appendix for an example of this using
cluster NGC 6397).

The goodness of fit criteria assumes that age and metallicity are uncorrelated, which we know
to be false.  However, as an iterative procedure, we can test each age and metallicity loop
separately and find a solution that converges in both parameters.  An example of the RMS space
for NGC 1904 is shown in Figure 4.  Each contour represents a normalized sum of the RMS values
for the age and metallicity loops per PC pair (in this example, age was iterated by 0.1 Gyrs
and [Fe/H] by 0.005 dex).  A minimum is found at accepted solution of 11.9 Gyrs and
[Fe/H]$=-1.55$ with a secondary minimum at 11.6 Gyrs.  While secondary minimums are common in
the sample, they were always located with 0.5 Gyrs of the primary minimum, i.e. within the
errors as given by observational error on the colors.  Also of interest is the direction of
the `valley' of low RMS which tracks from high metallicity, younger age to lower metallicities,
older age.  This is the age-metallicity degeneracy and the slope of the `valley' is $\delta$
log(age)/$\delta$($Z$)$=-1$ as predicted by Worthey (1999) for a near-blue filter system.  The
difference here is that a clear minimum can be determined (and the full range given by the
Schulz \etal models are searched), although any error in either age or metallicity will reflect
into the other with the slope given above (i.e. along the `valley').

\begin{figure}
\centering
\includegraphics[width=16cm]{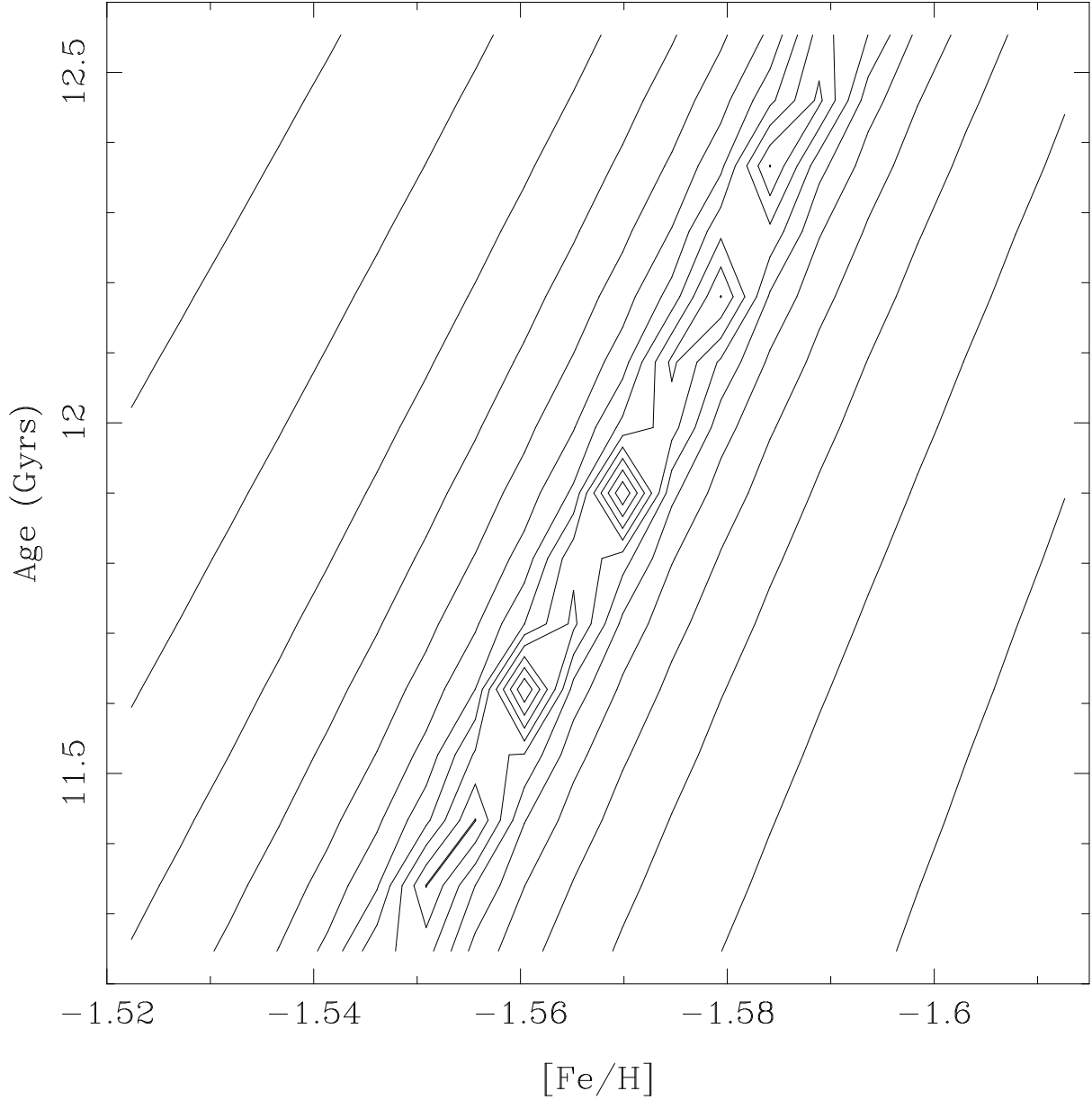}
\caption{The RMS contours of the iterative solutions for the cluster NGC 1904.  The lowest
contour represents an RMS of 0.001, increasing by a factor of three each contour.  The primary
minimum is found at 11.9 Gyrs and [Fe/H]$=-1.55$ with a secondary minimum at 11.6 Gyrs.  No
other minimums were found in the parameter space given by the Schulz \etal models (although
note that we only consider models older than 3 Gyrs).  }
\end{figure}

Inspection of the example in the Appendix demonstrates that the RMS minimums are shallower
for age loops compared to metallicity loops.  This confirms that our chosen filter system is
more sensitive to metallicity than age and, therefore, the uncertainty associated with age
will be greater than that of metallicity.  Examination of the RMS contours of the globular
clusters in the sample finds that the formal errors on the solutions are 0.4 Gyrs in age and
0.01 dex in [Fe/H], which is less than the error expected from observational uncertainties.
Hence, we conclude that the fitting process does not represent the dominant source of
uncertainty to our method and that the age-metallicity degeneracy is broken within the accuracy
of the data.  We note, however, the range of models we use to construct the PC pairs limits
this statement.  In particular, we do not examine models with ages less than 3 Gyrs as these
represent unrealistic ages when considering old stellar populations.  Thus, age-metallicity
degeneracy may once again come into play for very young populations or ones with super-solar
metallicities and we stress this caveat in any future usage of this technique.

To evaluate the accuracy of the entire technique, including the ability of the Schulz \etal
models to reproduce the integrated colors of globular clusters, we return to Figures 2 and 3,
the comparison between the PC solutions and the values for age and metallicity as determined by
CMD fitting from Salaris \& Weiss (2002).  Representative error bars are shown in the bottom
right hand portion of each diagram.  The error in the y-axis is taken from the mean error in
Salaris \& Weiss.  The error in the x-axis represents the values for the limits on our
technique discussed above.  Both age and metallicity are recovered through the narrow band
colors using our PC technique and the difference from the one-to-one line and a least-squares fit
is negligible.  Since the formal error for our solutions is less than the error associated with
CMD dating, it is tempting to claim that integrated colors are better predictors of age and
metallicity.  However, it is more likely that the interplay between the data in CMD's for
globular clusters and stellar evolutionary models is also present in our technique and the
Schulz \etal SSP models.  Thus, we adopt the scatter in Figures 2 and 3 as the real uncertainty
to our technique.  In Figure 2, the scatter around mean metallicity is approximately 0.2 dex,
greater than the internal error to our technique or the calibrating cluster metallicities.   In
Figure 3, the scatter around mean age is 0.5 Gyrs limited by the calibrating age from CMD fits
that have an error of $\pm$1 Gyr.  This appears to be an accuracy limit for globular clusters
(probably any type of SSP) until a set of better determined ages is found to test against our
method.

\section{METALLICITY CALIBRATION}

In a previous paper (Rakos \etal 2001), we have demonstrated a very tight correlation 
between $vz-yz$ color and the published metallicities for globular clusters deriving the 
relation:

\begin{equation}
[Fe/H] = (2.57 \pm 0.13)(vz-yz) - (2.20 \pm 0.04)
\end{equation}

\noindent where the scatter of individual values corresponds to the probable errors published
in the literature.  Examining the relation (1) for principal components, we find, as expected,
that the calculated metallicity value will be influenced, in a lesser manner, by the other two
color indices.  From an analysis of computed [Fe/H] values ([Fe/H]$_{ph}$, see Table 2), we find a
very similar expression (see Figure 5):

\begin{equation}
[Fe/H]_{ph} = (3.034 \pm 0.177)(vz-yz) - (2.305 \pm 0.049)
\end{equation}

Note that the fairly good correlation between $vz-yz$ and metallicity is due to the small value
for the coefficient of age in equation (1) plus the fact that the globular clusters used herein
have a small range of age (8 to 13 Gyrs).  Thus, this `fast' calibration from color to [Fe/H]
is only relevant for old stellar populations.  Also shown in Figure 5 is the relationship for
$bz-yz$, which is clearly not linear but does follow the predicted trend of bluer continuum
colors for lower metallicity clusters.  Since the calculated [Fe/H]$_{ph}$ values are related
to the real metallicity values (see Figure 2) by the following linear regression:

\begin{equation}
<Fe/H> = (0.876 \pm 0.065) [Fe/H]_{ph} - (0.174 \pm 0.104)
\end{equation}

\noindent where $<$Fe/H$>$ is mean value of [Fe/H] from Salaris \& Weiss.  Substituting for
[Fe/H]$_{ph}$ gives:

\begin{equation}
<Fe/H> = (2.658 \pm 0.089)(vz-yz) - (2.193 \pm 0.052)
\end{equation}

\noindent which is nearly identical to equation (3).  We can use this solution as a quick
estimation of metallicity with the knowledge of only one color index, $vz-yz$, since
$vz-yz$ is the most sensitive indicator of metallicity due to its measurement of an
absorption line rich region of the spectrum.  To achieve a more accurate estimate of metal
abundance, we would need the mean effective temperature of the stellar population in order to
calculate the abundance directly from the strength of absorption lines.  This implies that the
relationship between $vz-yz$ is only true for the systems composed of stellar populations of
identical temperatures.  Therefore, in the absence of temperature information, we combine all
three colors as given by PC1 resulting in:

\begin{equation}
X=0.480(uz-vz)+0.506(bz-yz)+0.511(vz-yz)
\end{equation}
 
\noindent and derive the approximation

\begin{equation}
[Fe/H] = -4.24 + 6.47X - 2.29X^2
\end{equation}

\noindent which is plotted in Figure 6 for visual comparison.  This relation delivers the same
values for the metallicity as the principal component analysis for globular clusters and SSP
models.  In addition, we can apply the same PC method outlined above on the multi-color data we
have acquired for dwarf and bright ellipticals from our Coma and field samples (Odell,
Schombert \& Rakos 2002).  The resulting [Fe/H] values for this sample of galaxies are also
shown in Figure 6, with the caveat here being that the galaxies are assumed to be SSP's like
globular clusters.  While galaxies follow the same trend as globulars in Figure 6, the
relationship is not linear.  We interpret this as confirmation that ellipticals are not SSP's,
but rather the summation of many SSP's to form a composite stellar population.  However, the
low order difference between globulars and ellipticals signals that the SSP assumption is an
adequate approximation for many needs and that the total stellar population in ellipticals must
be a simple form (such as a gaussian distribution) to produce the correlation seen in Figure 6
(see also Smolcic \etal 2004).  In fact, a simple 2nd order function can be fit to both the
globular cluster and elliptical data in Figure 6 (shown as the solid line) that empirically can
be used to determine the mean metallicity over 6 orders of magnitude in stellar mass.

\begin{figure}
\centering
\includegraphics[width=16cm]{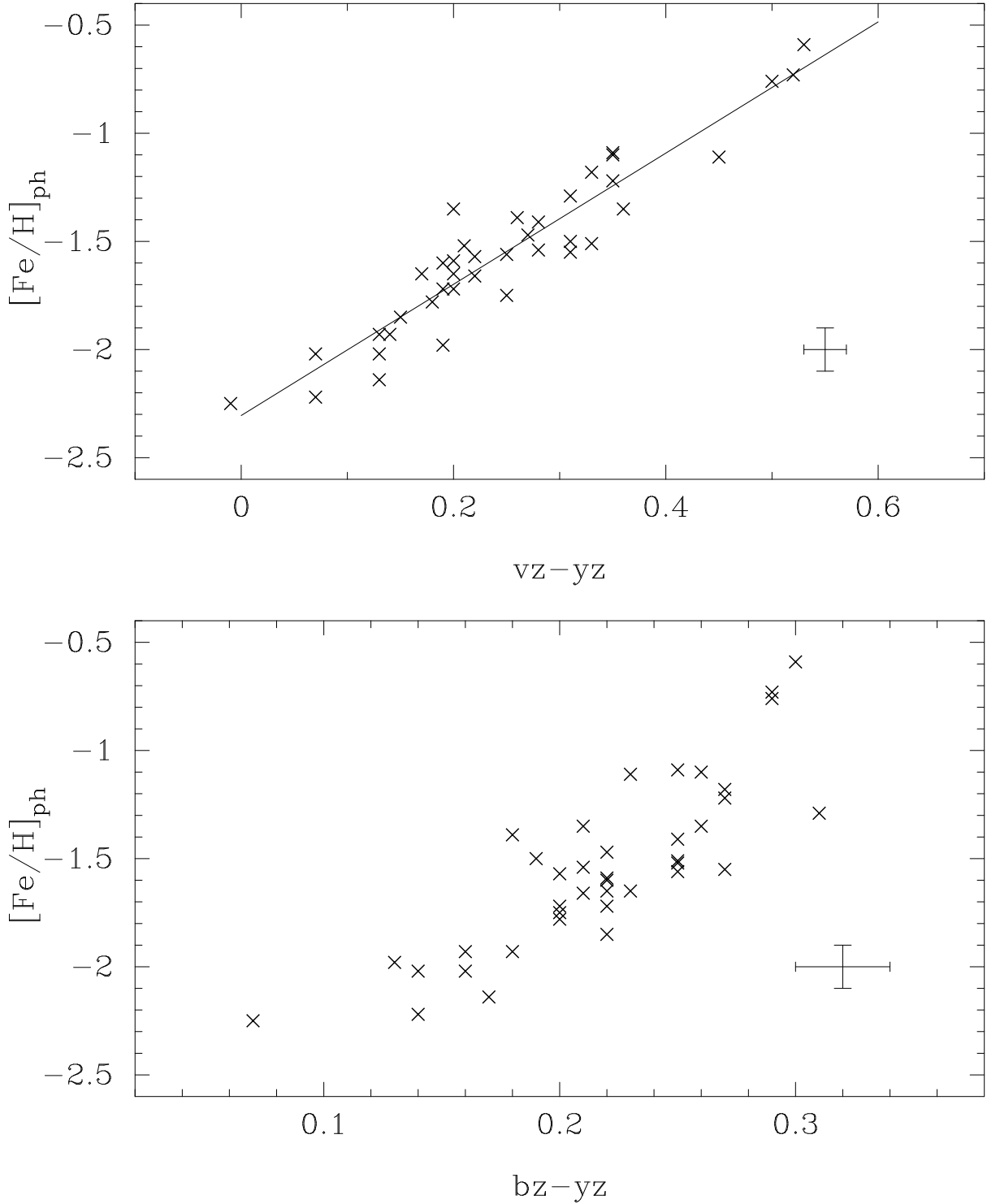}
\caption{The globular cluster $vz-yz$ and $bz-yz$ colors as a function of the calculated
[Fe/H]$_{ph}$.  The metallicity color, $vz-yz$, displays better linearity and lower scatter than
the continuum color, $bz-yz$.  An adequate measure of [Fe/H] can be obtained simply from the
$vz-yz$ color, if the population is older than 3 Gyrs.  The solid line is a least squares fit
to the $vz-yz$ data as quoted in the text.  }
\end{figure}

\begin{figure}
\centering
\includegraphics[width=12cm,angle=-90]{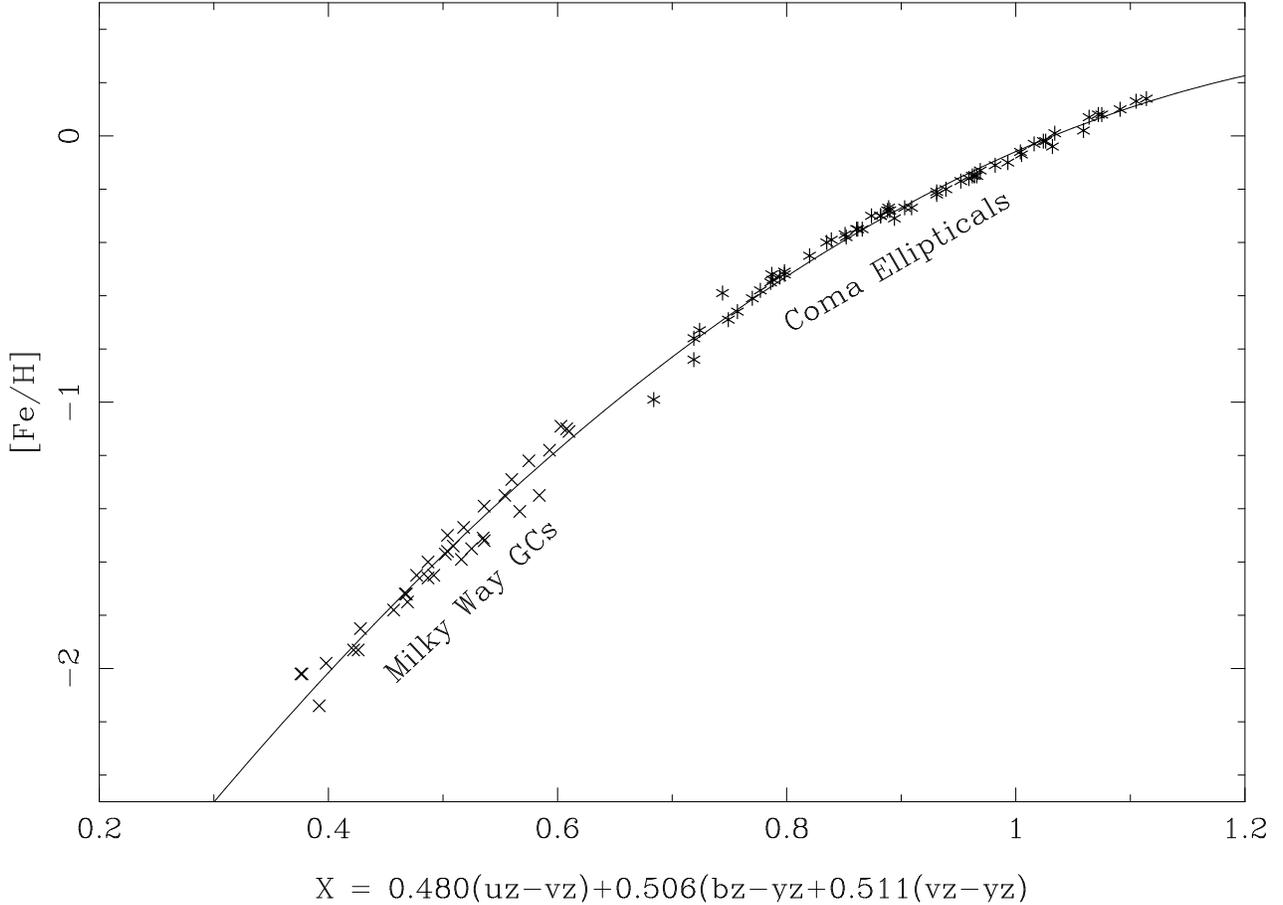}
\caption{The mean metallicity of the globular cluster sample as determined by the iterative
technique versus the multi-color term that incorporates all of the PC terms.  This relationship
can be compared to Figure 5 to demonstrate how the use of the full parameter space reduces the
scatter in [Fe/H].  Also shown are the iterative results for a sample of dwarf and giant
ellipticals in the Coma cluster with full $uvby$ photometry.  The relationship become less
linear for galaxies due to the fact that they are composed of a integrated population of SSP's.
The low scatter indicates that metallicity is the primary driver of galaxy color (see also Smolcic
\etal 2004).  }
\end{figure}

\section{AGE CALIBRATION}

Unfortunately, there is no similar simple solution for the estimation of age as there was for
metallicity.  In general, mean stellar age is correlated with metallicity since an older
population formed from low metallicity material, but this correlation is rather variable as
effected by the local environment and the local history of star formation.  And, of course,
this is the parameter we wish to determine in order to understand the galaxy evolution process.
Globular clusters in our Galaxy display such a difference between younger and older clusters as
divided by age on the order of 10 Gyrs where older clusters are more metal-poor.

The computed ages for measured globular clusters are listed in Table 2.  The recent compilation
and discussion of globular ages is found in Salaris \& Weiss (2002).  There are two sets of
ages with designation CG97 and ZW84 and, again, the mean is compared with our measurements
listed as $\tau_{ph}$ in Figure 3.  The correlation is better than expected and comparable to the
correlation between CG97 and ZW84 itself.  For comparison, we have at our disposal only 26
values from the paper of Salaris \& Weiss against our 40 photometric measurements.  We can
calculate the mean value of square differences between our estimation and the mean age of CG97
and ZW84, and these are plotted in Figure 7.  The scatter increases toward larger ages or for
metallicities below $-$1.7.  This implies a higher uncertainty for values extrapolated outside
the models we have used.  The realistic estimations of stellar population ages are, until now,
very scarce and, outside our Galaxy, practically impossible unless the underlying HR diagram
can be resolved (see Grebel 2004).

\begin{figure}
\centering
\includegraphics[width=16cm]{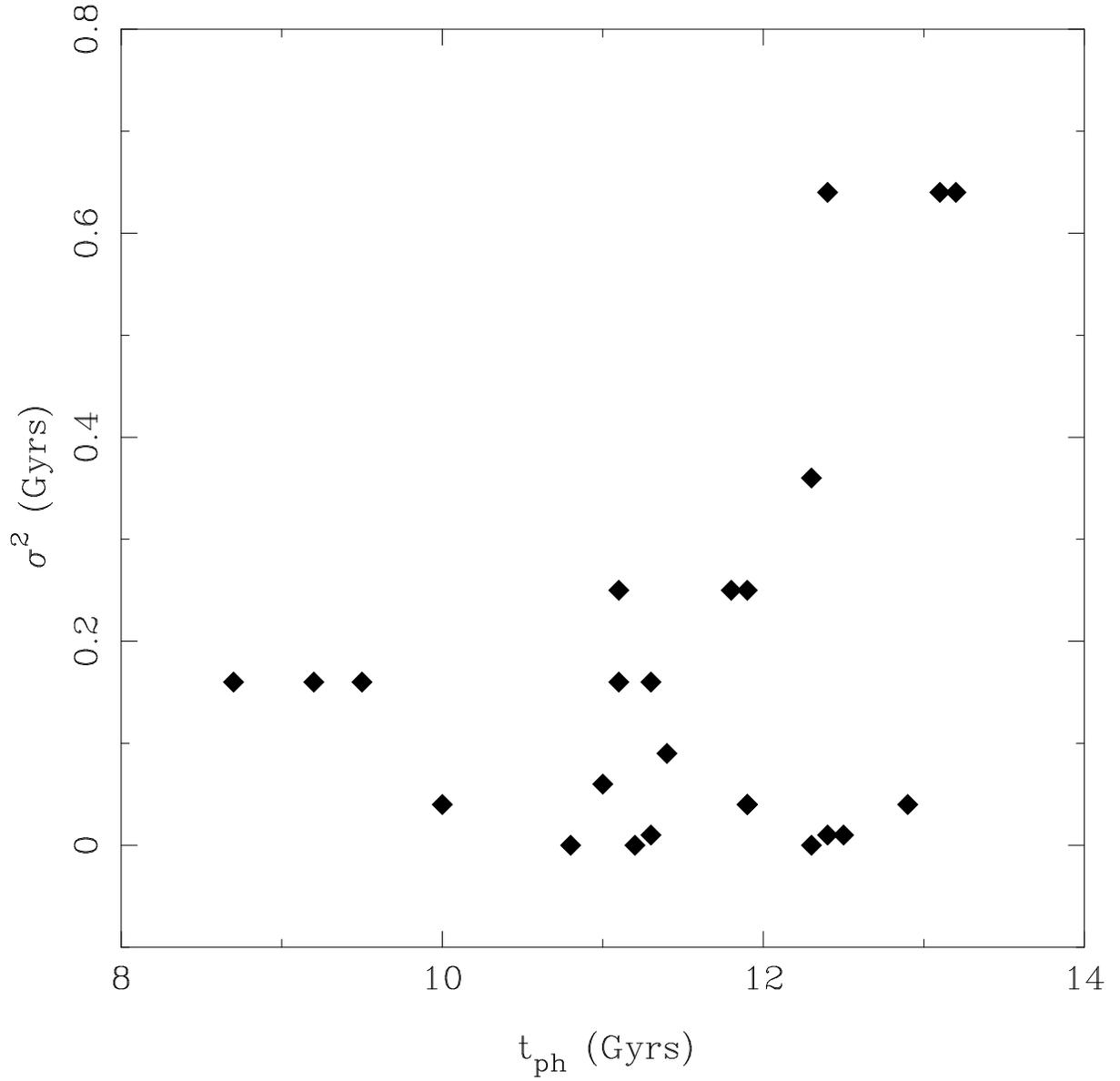}
\caption{The square of the differences between our iterative estimate of cluster age and the
true cluster age versus the calculated age, $\tau_{ph}$.  The difference increases with
decreasing age (and decreasing metallicity) implying higher uncertainty as one extrapolates
outside the model realm.  }
\end{figure}

An alternative approach would be to derive the metallicity of an object as shown in the
previous section, then apply a linear combination of [Fe/H] and all three narrow band colors.
For the globular cluster sample, this is shown in Figure 8, again using the PC values from
equation (2).  The resulting correlation is poor, although a least squares linear fit can be
made and is shown in Figure 8.  Much of the power in the fit is due to that fact that the
globular cluster sample has a fair correlation between metallicity and age, however, the
relative differences in age can be derived and the method shows promise as a method to separate
old and young galaxies in a relative sense, which can be a key test to hierarchical models of
galaxy formation (Rakos \& Schombert 2004).

\begin{figure}
\centering
\includegraphics[width=12cm,angle=-90]{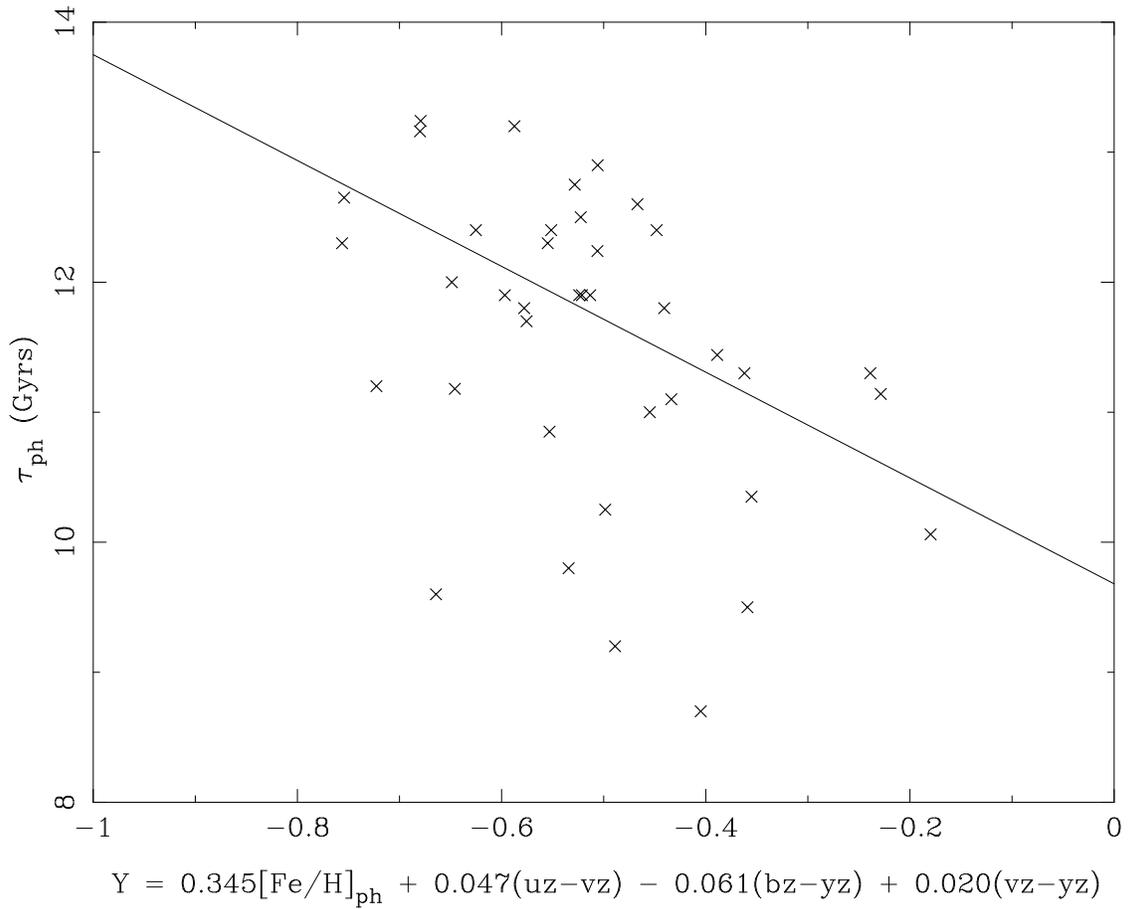}
\caption{The calculated cluster age, $\tau_{ph}$, by the iterative technique as a function
of the PC2 indices including [Fe/H].  While the correlation is not as clear as the metallicity
relationship, the general trend is discernible and will serve as a crude indicator of age for
more complex systems.}
\end{figure}

\section{SUMMARY}

In this paper we have revised the age and metallicity calibration for the Str\"omgren filter
system as applied to globular clusters and elliptical galaxies.  We have first investigated the
behavior of the color indices using the most recent SED models (Schulz \etal 2002) and applied
a principal component (PC) analysis on the SED models to determine the PC axis as well as the
range and limitations of the color system.  With this information, we have developed a
technique to derive age and metallicity for a simple stellar population by an iterative
calculation scheme.  To test this method, we have obtained new photometry of 40 globular
clusters with well-determined age and metallicity values and demonstrated that we can recover
their star formation history values using our iterative scheme.  Lastly, we have re-determined
the age and metallicity calibration for galaxies with new multi-color indices and tested the
calibration against a sample of field and cluster ellipticals that contain a variety of
spectroscopic and photometrically determined stellar population indicators.

Our next paper will be concentrate on the application of our method on dwarf and normal
ellipticals in clusters of galaxies.  However, we can already deduce from the low scatter in
[Fe/H] curves that the observed colors in ellipticals are primarily driven by metallicity, and
not age, effects.  While absolute age determination is problematic, relative age measurements
on the level of 1 Gyr can be extracted with sufficiently accurate photometry (errors less than
0.02 mags).

\appendix
\section{APPENDIX}

As an example to our technique to derive metallicity and age from the PC components of a SSP,
we demonstrate the calculation procedure for the globular cluster NGC 6397.  We have selected
NGC 6397 as an example since its [Fe/H] value ($-$1.88) is lower than the range of model values
and, thus, we must use an extrapolation allowing the metallicity loop to run over the border
given by the lowest model metallicity ($-$1.7).  Equation (6) gives an initial estimate for the
metallicity of NGC 6397 of [Fe/H]$=-$2.01, equation (8) predicts $-$2.13.

In Table 3, three pairs of age (12.4, 13.5 and 13.9 Gyrs) and metallicity ($-$1.966, $-$1.978
and $-$2.206) are selected to cover a broad range of age and metallicity at the boundary of
the published models.  Six of the iterated test values using equations (1) and (2) are shown
below each pair.  As each test converges on a unique PC1 and PC2 value, the range in age and
metallicity values becomes smaller.  This is measured by the mean values of each loop and the
sum of the squares of the deviations from mean. For the initial loop, these deviations are
large, as expected.  From further iterations we find the minimum to be between the first two
pairs.

Table 4 displays six more pairs for ages from 12.9 to 13.4 Gyr and metallicities from $-$2.014
to $-$2.034.  The top portion of the table displays the age results for six iterative searches
(metallicity loops are not shown).  There is little variation in age within each loop
represented by a nearly constant value for the sum of the squares of the deviations from mean
indicating a broad range of acceptable solutions for age.  Below the age loop values is a
single sequence of metallicity loops (age loop no. 4 in this case, although the variation from
loop to loop was small).  In the metallicity loop there is a detectable trend in the sum of the
deviations.  We can plot the selected pair values as a function of sum of the squares of the
deviations for the metallicity (Figure 9) to demonstrate the final result, a parabola with a
clear minimum that displays probable errors of about $\pm$0.03 dex in metallicity and $\pm$0.2
Gyr in age.

\begin{figure}
\centering
\includegraphics[width=12cm,angle=-90]{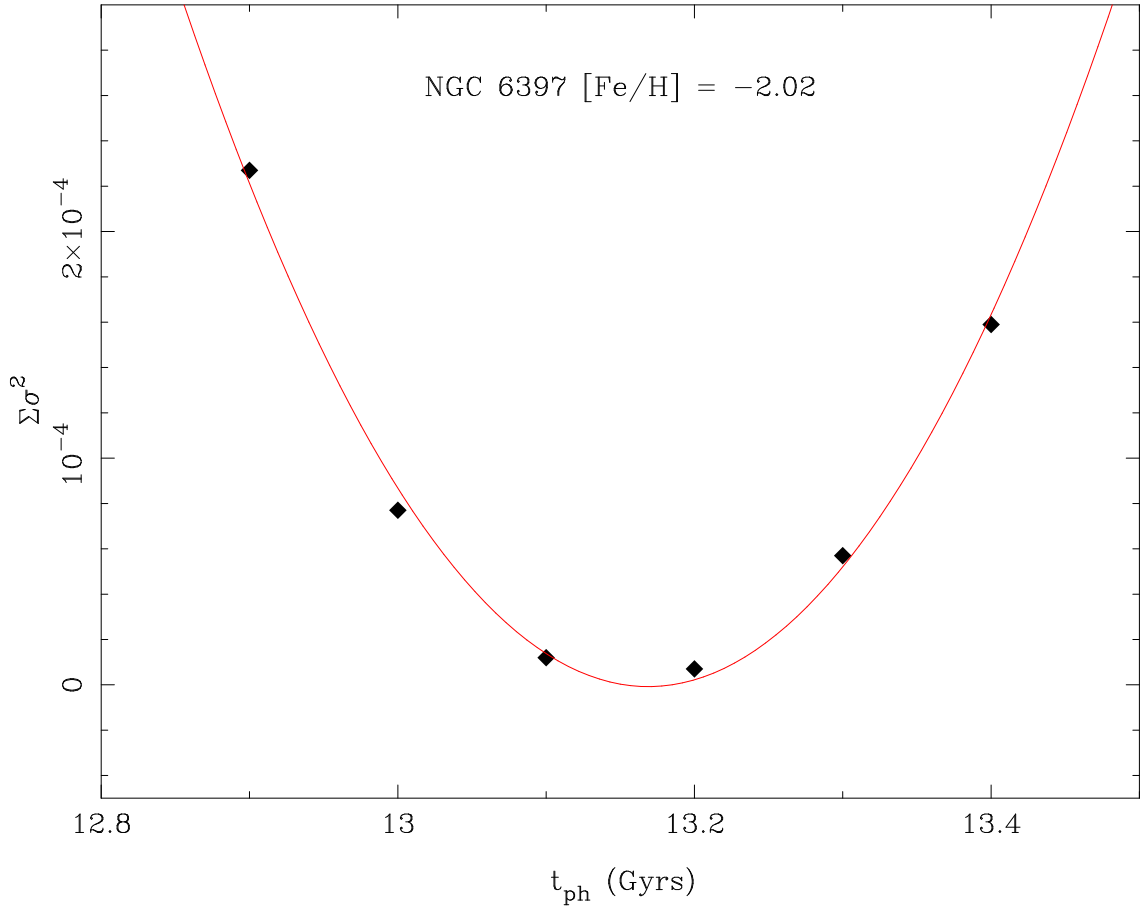}
\caption{A visual example of the iterative difference errors for six [Fe/H] values in the
globular cluster NGC 6397 data (see Table 4) where the sum of the squares of the deviations
from the mean is plotted versus the iteration pairs of age and metallicity.  The solid line is
a 2nd order fit to the errors displaying a clear minimum at age/metallicity pair 13.2 Gyrs and
[Fe/H]=$-$2.026.}
\end{figure}

While the solutions are older in age and lower in metallicity than the CMD fits from Salaris \&
Weiss, this cluster is an extreme example and outside the models shown in Figure 1, yet still
converges to a solution.  In addition, we note that Pasquini \etal (2004) find an age for NGC
6397 that is 0.3 Gyrs older than the Galactic disk based on beryllium measurements.
Str\"omgren photometry of the Galactic disk derives an estimate for its age between 13.2 and
13.5 Gyrs (Twarog 1980), which would make our age determination of 13.2 Gyrs for NGC 6397
exactly in line with the beryllium value.

\acknowledgements

The authors wish to thank the directors and staff of CTIO for granting time for this project.
Financial support from  Austrian Fonds zur Foerderung der Wissenschaftlichen Forschung and NSF
AST-0307508 is gratefully acknowledged.  We also wish to thank the referee, James Rose, for his
excellent suggestions to improve the presentation of our technique.

\begin{deluxetable}{crcrccc}
\tablecolumns{7}
\small
\tablewidth{0pt}
\tablecaption{SED Models}
\tablehead{
\colhead{PC1} &
\colhead{PC2} &
\colhead{[Fe/H]} &
\colhead{$\tau^a$} &
\colhead{$uz-vz$} &
\colhead{$bz-yz$} &
\colhead{$vz-yz$} \nl
}
\startdata
   0.160& -3.439 &-1.7& 3.08& 0.725 & 0.117 & 0.028
 \nl
   0.345& -4.357 &-1.7& 4.06& 0.691 & 0.137 & 0.066
 \nl
   0.710& -6.193 &-1.7& 6.02& 0.646 & 0.167 & 0.123
 \nl
   1.080& -8.158 &-1.7& 8.12& 0.620 & 0.179 & 0.140
 \nl
   1.428& -9.992 &-1.7&10.08& 0.599 & 0.190 & 0.159
 \nl
   1.779&-11.824 &-1.7&12.04& 0.606 & 0.192 & 0.165
 \nl
   2.139&-13.656 &-1.7&14.00& 0.623 & 0.199 & 0.177
 \nl
   0.820& -3.094 &-0.7& 3.08& 0.719 & 0.212 & 0.309
 \nl
   0.995& -4.011 &-0.7& 4.06& 0.716 & 0.215 & 0.317
 \nl
   1.367& -5.844 &-0.7& 6.02& 0.713 & 0.232 & 0.360
 \nl
   1.792& -7.807 &-0.7& 8.12& 0.727 & 0.258 & 0.432
 \nl
   2.143& -9.639 &-0.7&10.08& 0.731 & 0.260 & 0.442
 \nl
   2.531&-11.470 &-0.7&12.04& 0.760 & 0.275 & 0.489
 \nl
   2.867&-13.303 &-0.7&14.00& 0.759 & 0.272 & 0.479
 \nl
   1.003& -2.990 &-0.4& 3.08& 0.739 & 0.230 & 0.354
 \nl
   1.175& -3.907 &-0.4& 4.06& 0.718 & 0.231 & 0.375
 \nl
   1.572& -5.739 &-0.4& 6.02& 0.731 & 0.256 & 0.443
 \nl
   2.005& -7.701 &-0.4& 8.12& 0.768 & 0.279 & 0.513
 \nl
   2.372& -9.533 &-0.4&10.08& 0.785 & 0.287 & 0.537
 \nl
   2.726&-11.364 &-0.4&12.04& 0.801 & 0.288 & 0.543
 \nl
   3.113&-13.196 &-0.4&14.00& 0.829 & 0.303 & 0.587
 \nl
   1.263& -2.851 & 0.0& 3.08& 0.745 & 0.257 & 0.463
 \nl
   1.539& -3.765 & 0.0& 4.06& 0.805 & 0.293 & 0.576
 \nl
   1.869& -5.598 & 0.0& 6.02& 0.784 & 0.291 & 0.571
 \nl
   2.329& -7.557 & 0.0& 8.12& 0.872 & 0.312 & 0.648
 \nl
   2.693& -9.389 & 0.0&10.08& 0.886 & 0.320 & 0.669
 \nl
   3.109&-11.219 & 0.0&12.04& 0.945 & 0.340 & 0.736
 \nl
   3.501&-13.049 & 0.0&14.00& 0.996 & 0.349 & 0.776
 \nl
   1.597& -2.707 & 0.4& 3.08& 0.865 & 0.289 & 0.604
 \nl
   1.899& -3.619 & 0.4& 4.06& 0.949 & 0.332 & 0.737
 \nl
   2.275& -5.449 & 0.4& 6.02& 1.010 & 0.333 & 0.743
 \nl
   2.726& -7.409 & 0.4& 8.12& 1.062 & 0.356 & 0.836
 \nl
   3.126& -9.240 & 0.4&10.08& 1.108 & 0.375 & 0.884
 \nl
   3.484&-11.071 & 0.4&12.04& 1.130 & 0.375 & 0.894
 \nl
   3.907&-12.900 & 0.4&14.00& 1.212 & 0.392 & 0.955
 \nl
\enddata
\tablecomments{
$^a$ Age in Gyrs
}
\end{deluxetable}

\begin{deluxetable}{rcccccccc}
\tablecolumns{9}
\small
\tablewidth{0pt}
\tablecaption{Globular Cluster Data}
\tablehead{
\colhead{NGC} &
\colhead{$uz-vz$} &
\colhead{$bz-yz$} &
\colhead{$vz-yz$} &
\colhead{E(B-V)} &
\colhead{[Fe/H]$^a$} &
\colhead{$\tau^b$} &
\colhead{[Fe/H]$_{ph}$} &
\colhead{$\tau_{ph}^c$} \nl
}
\startdata
104  & 0.65 & 0.29 & 0.52 & 0.04 & $-$0.75 & 10.7 & $-$0.73 & 11.1 \nl
288  & 0.51 & 0.31 & 0.31 & 0.03 & $-$1.26 & 11.6 & $-$1.29 & 11.1 \nl
362  & 0.54 & 0.27 & 0.35 & 0.05 & $-$1.19 & 9.1 & $-$1.22 & 8.7 \nl
1261 & 0.56 & 0.22 & 0.27 & 0.01 & $-$1.25 & 8.8 & $-$1.47 & 9.2 \nl
1851 & 0.62 & 0.26 & 0.35 & 0.02 & $-$1.16 & 9.1 & $-$1.10 & 9.5 \nl
1904 & 0.60 & 0.20 & 0.22 & 0.01 & $-$1.54 & 12.1 & $-$1.57 & 11.9 \nl
2298 & 0.63 & 0.25 & 0.21 & 0.14 & $-$1.80 & 12.7 & $-$1.52 & 12.9 \nl
3201 & 0.60 & 0.27 & 0.33 & 0.23 & $-$1.45 & 11.7 & $-$1.18 & 11.4 \nl
4147 & 0.58 & 0.16 & 0.13 & 0.02 & $-$1.83 & -- & $-$1.93 & 11.2 \nl
4372 & 0.40 & 0.14 & 0.07 & 0.39 & $-$2.09 & -- & $-$2.22 & 12.7 \nl
4590 & 0.50 & 0.17 & 0.13 & 0.05 & $-$2.06 & 11.2 & $-$2.14 & 11.2 \nl
4833 & 0.54 & 0.18 & 0.14 & 0.32 & $-$1.80 & -- & $-$1.93 & 12.0 \nl
5024 & 0.50 & 0.14 & 0.13 & 0.02 & $-$1.99 & -- & $-$2.02 & 13.2 \nl
5139 & 0.53 & 0.22 & 0.20 & 0.12 & $-$1.62 & -- & $-$1.72 & 11.8 \nl
5272 & 0.54 & 0.21 & 0.28 & 0.01 & $-$1.52 & 11.7 & $-$1.54 & 11.9 \nl
5286 & 0.56 & 0.20 & 0.19 & 0.24 & $-$1.67 & -- & $-$1.72 & 11.7 \nl
5466 & 0.52 & 0.07 & $-$0.01 & 0.00 & $-$2.19 & 12.3 & $-$2.25 & 12.3 \nl
5634 & 0.49 & 0.13 & 0.19 & 0.05 & $-$1.88 & -- & $-$1.98 & 9.6 \nl
5897 & 0.58 & 0.22 & 0.20 & 0.09 & $-$1.82 & 12.3 & $-$1.65 & 12.4 \nl
5904 & 0.65 & 0.18 & 0.26 & 0.03 & $-$1.26 & 11.3 & $-$1.39 & 11.0 \nl
5986 & 0.58 & 0.22 & 0.19 & 0.28 & $-$1.58 & -- & $-$1.60 & 9.8 \nl
6101 & 0.57 & 0.23 & 0.17 & 0.05 & $-$1.80 & 10.8 & $-$1.65 & 10.9 \nl
6171 & 0.55 & 0.23 & 0.45 & 0.33 & $-$1.03 & 11.7 & $-$1.11 & 11.3 \nl
6205 & 0.50 & 0.20 & 0.25 & 0.02 & $-$1.50 & 12.4 & $-$1.75 & 13.2 \nl
6218 & 0.44 & 0.17 & 0.28 & 0.19 & -- & 12.6 & -- & -- \nl
6229 & 0.52 & 0.19 & 0.31 & 0.01 & $-$1.43 & -- & $-$1.50 & 10.3 \nl
6352 & 0.67 & 0.30 & 0.53 & 0.21 & $-$0.63 & 9.8 & $-$0.59 & 10.1 \nl
6397 & 0.54 & 0.16 & 0.07 & 0.18 & $-$1.88 & 12.3 & $-$2.02 & 13.2 \nl
6584 & 0.56 & 0.21 & 0.22 & 0.10 & $-$1.44 & 11.7 & $-$1.66 & 12.3 \nl
6652 & 0.66 & 0.29 & 0.50 & 0.09 & $-$0.89 & 11.4 & $-$0.76 & 11.3 \nl
6656 & 0.48 & 0.27 & 0.31 & 0.34 & $-$1.60 & 12.4 & $-$1.55 & 12.5 \nl
6681 & 0.52 & 0.25 & 0.25 & 0.07 & $-$1.46 & 11.7 & $-$1.56 & 11.9 \nl
6715 & 0.50 & 0.25 & 0.33 & 0.15 & $-$1.58 & -- & $-$1.51 & 12.2 \nl
6723 & 0.56 & 0.26 & 0.36 & 0.05 & $-$1.07 & 11.6 & $-$1.35 & 12.4 \nl
6752 & 0.55 & 0.20 & 0.18 & 0.04 & $-$1.45 & 12.4 & $-$1.78 & 11.9 \nl
6809 & 0.72 & 0.21 & 0.20 & 0.08 & $-$1.79 & 12.3 & $-$1.35 & 11.8 \nl
6864 & 0.62 & 0.25 & 0.35 & 0.16 & $-$1.16 & -- & $-$1.09 & 10.4 \nl
6981 & 0.62 & 0.25 & 0.28 & 0.05 & $-$1.40 & -- & $-$1.41 & 12.6 \nl
7078 & 0.50 & 0.22 & 0.15 & 0.10 & $-$2.14 & -- & $-$1.85 & 12.4 \nl
7089 & 0.63 & 0.22 & 0.20 & 0.06 & $-$1.62 & -- & $-$1.59 & 12.8 \nl
\enddata
\tablecomments{
$^a$ Mean value from Harris (1996) and Salaris \& Weiss (2002)
$^b$ Salaris \& Weiss (2002), age in Gyrs
$^c$ Age in Gyrs
}
\end{deluxetable}

\begin{deluxetable}{rccc}
\tablecolumns{7}
\small
\tablewidth{0pt}
\tablecaption{NGC 6397 Initial Loop}
\tablehead{
\colhead{Age (Gyrs)/[Fe/H]} &
\colhead{12.4/$-$1.966} &
\colhead{13.5/$-$1.978} &
\colhead{13.9/$-$2.206} \nl
}
\startdata
Test no: 1 & 12.42/$-$1.944 
           & 13.53/$-$1.935 
           & 13.88/$-$2.322 \nl
         2 & 12.45/$-$1.904 
           & 13.57/$-$1.861 
           & 13.84/$-$2.528 \nl
         3 & 12.49/$-$1.834 
           & 13.63/$-$1.733 
           & 13.75/$-$2.905 \nl
         4 & 12.54/$-$1.713 
           & 13.71/$-$1.713 
           & 13.57/$-$3.626 \nl
         5 & 12.62/$-$1.504 
           & 13.85/$-$1.163 
           & 13.19/$-$5.123 \nl
         6 & 12.75/$-$1.155 
           & 14.05/$-$0.601 
           & 12.31/$-$8.850 \nl
Mean       & 12.54/$-$1.676 
           & 13.72/$-$1.469 
           & 13.42/$-$4.226 \nl
Sum $\sigma^2$  & 0.075/0.4512 
                & 0.193/1.2900 
                & 1.804/30.799 \nl
\enddata
\end{deluxetable}

\begin{deluxetable}{rcccccc}
\tablecolumns{7}
\small
\tablewidth{0pt}
\tablecaption{NGC 6397 Age/[Fe/H] Loops}
\tablehead{
\colhead{Selected Age} & 
\colhead{12.9/} &
\colhead{13.0/} &
\colhead{13.1/} &
\colhead{13.2/} &
\colhead{13.3/} &
\colhead{13.4/} \nl
\colhead{and [Fe/H]} & 
\colhead{$-$2.014} &
\colhead{$-$2.018} &
\colhead{$-$2.022} &
\colhead{$-$2.026} &
\colhead{$-$2.030} &
\colhead{$-$2.034} \nl
}
\startdata
Age Loop no: 1 & 12.92 & 13.02 & 13.12 & 13.22 & 13.32 & 13.42 \nl
         2 & 12.93 & 13.04 & 13.14 & 13.24 & 13.34 & 13.44 \nl
         3 & 12.95 & 13.05 & 13.15 & 13.25 & 13.35 & 13.45 \nl
         4 & 12.97 & 13.07 & 13.17 & 13.27 & 13.37 & 13.47 \nl
         5 & 12.99 & 13.09 & 13.19 & 13.29 & 13.39 & 13.49 \nl
         6 & 13.01 & 13.11 & 13.21 & 13.31 & 13.41 & 13.51 \nl
Mean       & 12.96 & 13.06 & 13.16 & 13.26 & 13.36 & 13.46 \nl
Sum $\sigma^2$ ($\times 10^{-4}$) & 61 & 55 & 55 & 55 & 55 & 55 \nl
&&&&&&\nl
[Fe/H] Loop no: 1 & $-$2.014 & $-$2.018 & $-$2.022 & $-$2.027 & $-$2.031 & $-$2.035 \nl
         2 & $-$2.014 & $-$2.018 & $-$2.023 & $-$2.027 & $-$2.032 & $-$2.036 \nl
         3 & $-$2.013 & $-$2.018 & $-$2.023 & $-$2.028 & $-$2.033 & $-$2.038 \nl
         4 & $-$2.010 & $-$2.016 & $-$2.023 & $-$2.029 & $-$2.035 & $-$2.040 \nl
         5 & $-$2.005 & $-$2.014 & $-$2.022 & $-$2.029 & $-$2.037 & $-$2.044 \nl
         6 & $-$1.997 & $-$2.008 & $-$2.019 & $-$2.030 & $-$2.040 & $-$2.050 \nl
Mean       & $-$2.009 & $-$2.015 & $-$2.022 & $-$2.028 & $-$2.035 & $-$2.041 \nl
Sum $\sigma^2$ ($\times 10^{-6}$)  &  227 &  77 &  12  &  7  &  57  & 159 \nl
\enddata
\end{deluxetable}

\end{document}